\documentclass[preprint,3p,10pt]{elsarticle}

\usepackage{graphicx}
\usepackage{subcaption}
\usepackage{color}
\usepackage{mwe}
\usepackage{wrapfig}

\usepackage{float} 
\usepackage{fullpage}
\usepackage{adjustbox}
\usepackage{amsfonts}
\usepackage{amsmath}
\usepackage{mathtools}
\usepackage{multicol}
\usepackage{amssymb}
\usepackage{amsbsy}
\usepackage{amsthm}
\usepackage{epsfig}
\usepackage{graphicx}
\usepackage{colordvi}
\usepackage{graphics}
\usepackage{color}
\usepackage{bm}
\usepackage{url}
\usepackage{algorithm}
\usepackage{algorithmic}
\usepackage{verbatim} 
\usepackage{braket}
\biboptions{sort&compress}

\usepackage{accents}

\begin{document}
\begin{frontmatter}

\title{Simulating the spread of COVID-19 \textsl{via} a spatially-resolved susceptible-exposed-infected-recovered-deceased (SEIRD) model with heterogeneous diffusion}

\author[dicar]
{Alex Viguerie}
\author[oden]{Guillermo Lorenzo}

\author[dicar]{Ferdinando Auricchio}
\author[rwth]{Davide Baroli}
\author[oden]{Thomas J.R. Hughes}
\author[dicar]{Alessia Patton}
\author[dicar]{Alessandro Reali}

\author[tey,oden]{Thomas E. Yankeelov}
\author[emory,emoryCS]{Alessandro Veneziani\corref{cor1}}
\ead{avenez2@emory.edu}

\cortext[cor1]{Corresponding author}
\address[dicar]{Dipartimento di Ingegneria Civile ed Architettura, Universit\`a di Pavia, Via Ferrata 3, Pavia, PV 27100, Italy}
\address[emory]{Department of Mathematics, Emory University, 400 Dowman Drive, Atlanta, GA 30322
, USA}
\address[emoryCS]{Department of Computer Science, Emory University, 400 Dowman Drive, Atlanta, GA 30322
, USA}
\address[tey]{Departments of Biomedical Engineering, Diagnostic Medicine, and Oncology, Livestrong Cancer Institutes, The University of Texas at Austin, 107 W. Dean Keeton St., Austin, TX 78712, USA}
\address[oden]{Oden Institute for Computational Engineering and Sciences, The University of Texas at Austin, 201 E. 24th Street, Austin, TX 78712-1229, USA}
\address[rwth]{ Aachen Institute for Advanced Study in Computational Engineering Science (AICES), RWTH Aachen University, Schinkelstra{\ss}e 2, 52062 Aachen, Germany}
\begin{keyword}
Partial differential equations, mathematical epidemiology, compartmental models, COVID-19, mathematical modeling, mathematical biology
\end{keyword}

\begin{abstract}
We present an early version of a Susceptible-Exposed-Infected-Recovered-Deceased (SEIRD) mathematical model based on partial differential equations coupled with a heterogeneous diffusion model. The model describes the spatio-temporal spread of the COVID-19 pandemic, and aims to capture dynamics also based on human habits and geographical features. To test the model, we compare the outputs generated by a finite-element solver with measured data over the Italian region of Lombardy, which has been heavily impacted by this crisis between February and April 2020. Our results show a strong qualitative agreement between the simulated forecast of the spatio-temporal COVID-19 spread in Lombardy and epidemiological data collected at the municipality level. Additional simulations exploring alternative scenarios for the relaxation of lockdown restrictions suggest that reopening strategies should account for local population densities and
the specific dynamics of the contagion. Thus, we argue that data-driven simulations of our model
could ultimately inform health authorities to design effective pandemic-arresting measures and
anticipate the geographical allocation of crucial medical resources.  \end{abstract}
\end{frontmatter}

\section{Introduction}\label{sec:intro}

The outbreak of COVID-19 in 2020 has caused widespread disruption throughout the world, leading to substantial damage in terms of both human lives and economic cost. To arrest the spread of the disease, governments have enacted unprecedented measures, including quarantines, curfews, lockdowns, and suspension of travel. The wide-reaching ramifications of such measures, deemed by many experts as necessary, are driven in part by a lack of clear information about the spatio-temporal spread of COVID-19. Indeed, the absence of reliable data regarding disease transmission has necessarily led to cautious responses. These recent events, which have required important decisions based on forecasts, have demonstrated more than ever the need for reliable tools intended to model the spread of COVID-19 and other infectious diseases \cite{remuzzi2020covid}. A particularly urgent need is the geo-localization of outbreaks, as this may allow a more effective allocation of medical resources. 
 
Several notable models of this outbreak have been presented; indeed, at the time of this writing there are over 1,000 COVID-19 articles on MedRXiv, many of which address the modeling of disease spread.
Some models aim at offering specific evaluations of policy responses based on the implementation of different social distancing measures with combined compartmental and empirical approaches \cite{Ferguson2020, Gatto202004978, GBB2020}. Rather than adopting a deterministic, mechanism-based model, Zhang \textit{et al.} employed a statistical approach to analyze the spatio-temporal dynamics of COVID-19 \cite{Zhang2020.03.23.20034058}.
In Gatto \textit{et al.} a combined statistical and compartmental approach was employed, in which spatial dependence is addressed by dividing the region of interest (Italy) into local communities connected by a network structure \cite{Gatto202004978}.

Here, we propose an alternative approach, using a partial-differential-equation (PDE) model designed 
to capture the continuous spatio-temporal dynamics of COVID-19. We leverage a compartmental SEIRD (\textit{susceptible, exposed, infected, recovered, deceased}) model that incorporates the spatial spread of the disease with inhomogeneous diffusion terms
\cite{HLBV1994, KGV2013, K1996, CG2010}. 
The rationale is that the diffusion operator, properly tuned to account for local natural 
or social inhomogeneities (e.g., mountains, rivers, highways) may describe 
the local movement of the different populations in a deterministic way, as the limit of a Brownian 
motion \cite{SSalsa}. This is critical to accurately account for information relevant to the outbreak dynamics, such as local population densities, which vary in space and time. While a mathematical description of non-local dynamics 
is still possible in terms of fractional differential operators \cite{khan2020modeling},
we postpone this approach to a follow-up of the present work. 
\par Hence, our modeling approach is more appropriate for the local dynamics on mesoscales, such as regions within Italy. Thus, to evaluate the model efficacy, we run a simulation study of the COVID-19 outbreak in the Italian region of Lombardy, which has been severely impacted by the COVID-19 crisis between February and April of 2020 and for which the necessary data was available. Also, the high density of Lombardy's population and transportation network is specifically suitable to our modeling approach. Our simulations show a remarkable qualitative agreement with the reported epidemiological data. We further explore various reopening scenarios, obtaining contrasting results that highlight the importance of considering local population densities and contagion dynamics.

The paper outline is as follows. In Section~\ref{model}, we describe the SEIRD model. Then, Section~\ref{numerics} addresses the numerical implementation of our model and Section~\ref{results} presents the results of the simulation study in Lombardy. We conclude in Section~\ref{disc} by examining the shortcomings observed from our simulations and discussing the additional work required to improve model accuracy and practical relevance.


\section{Model}\label{model}
Let $\Omega \subset \mathbb{R}^2 $ be a simply connected domain of interest and $\lbrack 0,\,T\rbrack$ a generic time interval. We denote the densities of the \textit{susceptible}, \textit{exposed}, \textit{infected}, \textit{recovered} and \textit{deceased} populations as $s(\boldsymbol{x},t)$, $e(\boldsymbol{x},t)$, $i(\boldsymbol{x},t)$, $r(\boldsymbol{x},t)$, and $d(\boldsymbol{x},t)$  respectively. Also, let $n(\boldsymbol{x},t)$ denote the sum of the living population; i.e., $n(\boldsymbol{x},t)=s(\boldsymbol{x},t)+e(\boldsymbol{x},t)+i(\boldsymbol{x},t)+r(\boldsymbol{x},t)$. Then, our model is comprised of the following system of coupled PDEs over $\Omega \times \lbrack 0,\,T\rbrack$:
\begin{align}
\label{eq1} \partial_t s &= \alpha n - \left(1-A/n\right)\beta_i s i - \left(1-A/n\right)\beta_e s e - \mu s n + \nabla\cdot\left(n\, \nu_s \nabla s\right) \\
\label{eq2} \partial_t e &= \left(1-A/n\right)\beta_i si + \left(1-A/n\right)\beta_e se - \sigma e - \phi_e e - \mu en + \nabla\cdot\left(n\, \nu_e \nabla e \right) \\
\label{eq3} \partial_t i &= \sigma e - \phi_d \,i - \phi_r i - \mu i n + \nabla \cdot\left(n\, \nu_i \nabla i \right)\\
\label{eq4} \partial_t r &= \phi_r i + \phi_e e - \mu r n + \nabla \cdot\left(n\, \nu_r \nabla r \right) \\
\label{eq5} \partial_t d &= \phi_d\, i,	
\end{align}
where $\alpha$ is the birth rate, $\sigma$ is the inverse of the incubation period, $\phi_e$ is the asymptomatic recovery rate, $\phi_r$ is the infected recovery rate, $\phi_d$ is the infected mortality rate, $\beta_e$ is the asymptomatic contact rate, $\beta_i$ is the symptomatic contact rate, $\mu$ is the general (non-COVID-19) mortality rate, and $\nu_s$, $\nu_e$, $\nu_i$, and $\nu_r$ are diffusion parameters respectively corresponding to the different population groups. Each of these parameters may depend on time, space, or the model compartments. We also consider the Allee effect (depensation), characterized by the parameter $A$. In this particular setting, the Allee effect serves to model the tendency of outbreaks to cluster towards large population centers. Specific parameter selection as well as initial and boundary conditions are discussed in Section~\ref{numerics}. 

\begin{figure}[t]
\centering
  \includegraphics[width=\linewidth]{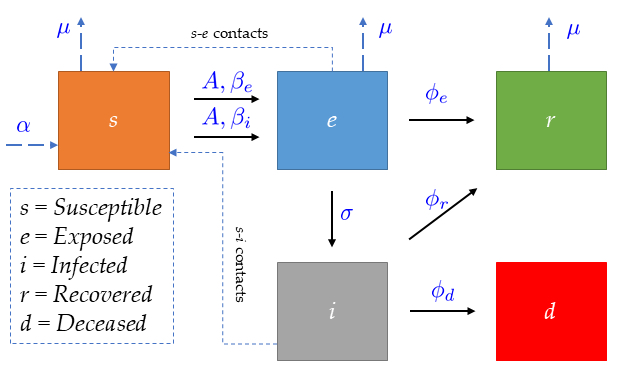}
 \caption{Flow chart describing the dynamics of contagion between the population subgroups considered in our model}\label{fig:flowChart}
\end{figure}

Fig.~\ref{fig:flowChart} shows the dynamics of contagion between the compartments in our model. We remark that our model accounts for asymptomatic transmission, which is considered a pivotal driver of the COVID-19 pandemic \cite{D2020,GBB2020,NLA2020}.
Eqs.~\eqref{eq1}--\eqref{eq2} show that exposed asymptomatic patients may transmit COVID-19 to susceptible individuals at contact rate $\beta_e$. This aligns with recent studies suggesting that patients may transmit COVID-19 almost immediately after exposure \cite{D2020, GBB2020,  NLA2020}. Additionally, Eqns.~\eqref{eq2} and \eqref{eq4} involve a fraction $\phi_e$ of exposed patients that do not develop symptoms and move directly into the recovered population. We also assume that recovered patients are immune, as we do not include any backflow from  Eq.~\eqref{eq4} to Eq.~\eqref{eq1}.
(We note that this is a current source of debate, but consistent with the existing literature for the time scale of months considered here \cite{LXY2020}). The spatial movement over a large population is 
described by an inhomogeneous random walk, which in the limit tends to a second order differential operator \cite{SSalsa}. The diffusivity coefficient is proportional to the population and can be locally adjusted to incorporate geographical or human-related inhomogeneities \cite{KGV2013}.

\section{Numerical implementation}\label{numerics}
We use a finite-element spatial discretization of the Italian region of Lombardy, consisting of an unstructured mesh containing 30,407 triangles (a mesh convergence analysis was first performed; data not shown). We use the backward-Euler method for time integration and solve each time step fully implicitly with a Picard iteration for stability. The resulting linear systems are solved by the GMRES algorithm using a Jacobi preconditioner.

\begin{figure}[t]
\centering
  \includegraphics[width=\linewidth]{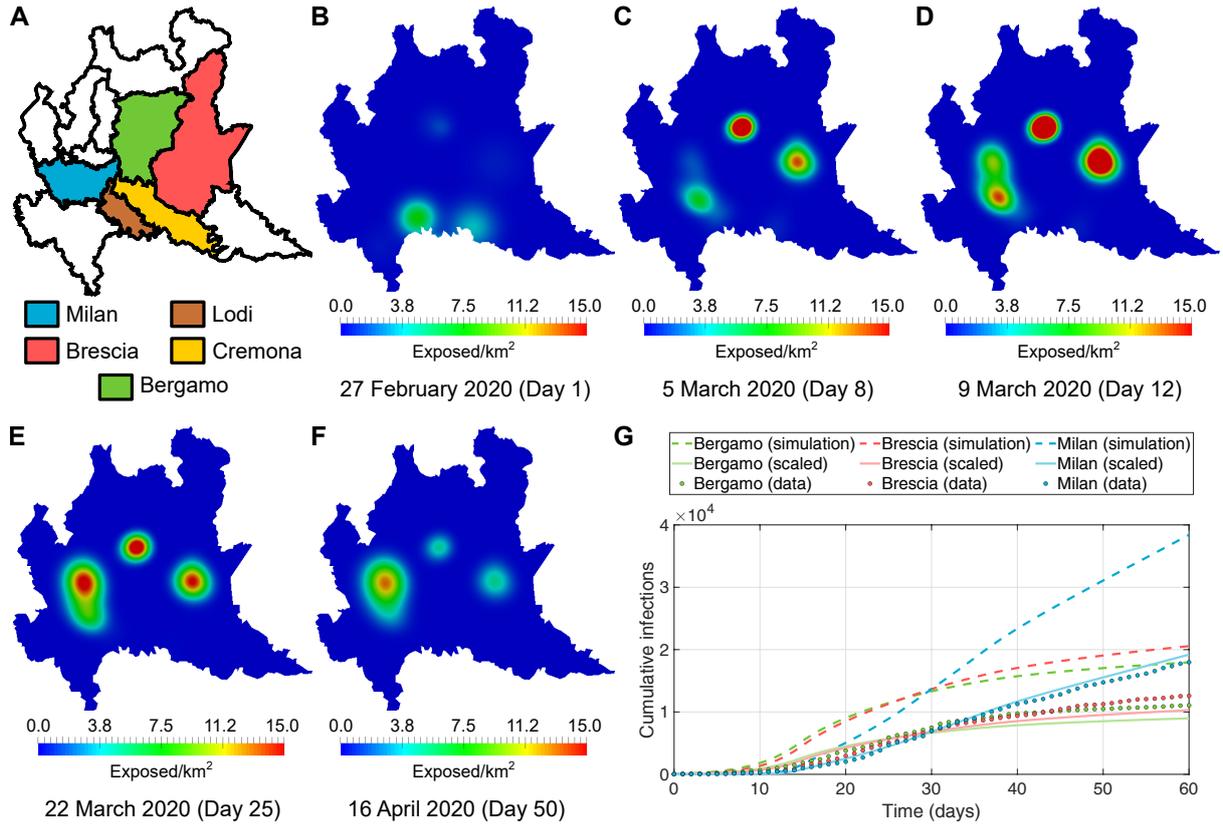}
 \caption{Model forecast of COVID-19 spread in Lombardy. (A) Main areas affected by the pandemic in Lombardy. (B) Initially, the main affected areas are Lodi and Cremona and, to lesser extent, Bergamo and Brescia.(C-E) Our model predicts increasing exposures in Bergamo and Brescia. The outbreak in Lodi soon moves north into the Milan metro area, where it further spreads despite the lockdown restrictions. (F) The model also predicts that governmental restrictions eventually succeed in reducing the exposure to the disease, which is faster in Brescia and Bergamo than in Milan. (G) Cumulative curves of infections according to reported data (dots) and simulations (dashed lines) for the three main areas of contagion: Bergamo, Brescia, and Milan. The model has been calibrated to match the data reported for the deceased subgroup, resulting in a forecast of a larger number of infections. To highlight the qualitative agreement of our simulations, we also show the numerical results scaled to match the order of magnitude of the reported infectious data (solid lines).}\label{fig:outbreakDevelopment}
\end{figure}

Initial conditions for the subpopulations $s$, $e$, $i$, $r$, and $d$ in the model are defined by means of Gaussian circular functions centered at the latitude and longitudinal coordinates of each municipality with 10,000 or more inhabitants, and weighted by the municipality's population size and geographic area. These initial conditions correspond to the data provided by Lab24 \cite{Lab24} from the date 27 February 2020, featuring a severe outbreak in the province of Lodi, and moderate numbers of exposed and infected individuals in the provinces of Bergamo, Brescia, and Cremona (see Fig.~\ref{fig:outbreakDevelopment}). We used homogeneous Neumann boundary conditions for simplicity, mimicking a complete isolation of the region. We acknowledge that this is likely unrealistic (e.g.  if lockdown orders are relaxed) and will be considered in future efforts.

We assume $\sigma$=$1/7$ day$^{-1}$, $\phi_r$=1/24 day$^{-1}$, $\phi_m$=1/160 day$^{-1}$, and $\phi_e$=1/6 day$^{-1}$. These values were based on available data from the literature regarding the mortality, incubation period, and recovery time for infected and asymptomatic patients \cite{BKW2020, Ferguson2020, GBB2020, NLA2020, D2020}. Additionally, we do not consider births or non-COVID-19 mortality (i.e., we set $\alpha=0$ and $\mu=0$, respectively), given the time scale of months in our simulations.

The remainder of the model parameters are estimated in a two-step approach. First, we fit a 0D SEIRD version of our model (i.e., consisting of a system of exclusively time-dependent ordinary differential equations and no diffusion terms) to match the temporal dynamics of the outbreak. Then, we iteratively refined these values by means of recursive simulations using Eqs.~\eqref{eq1}--\eqref{eq5} to match the spatio-temporal epidemiological data. We use the $R^2$ coefficient and the root mean squared error (RMSE) to assess the goodness-of-fit.

Given the uncertainty in the currently available COVID-19 data, we think that parameter estimation aiming at matching the dynamics of all model compartments is not viable. As not every member of the population is tested for infection and asymptomatic cases are known to exist in possibly large numbers, we think that the available data of infected cases might lead to unrealistic parameter fitting. Conversely, the data reported for COVID-19 deaths offer more reliability to calibrate the model parameters. Therefore, we pursue quantitative agreement in the deceased compartment (i.e., $d$), and qualitative agreement for the rest of the model subgroups (i.e., $s$,$e$,$i$,$r$). In Section 4, our results will focus on the model forecasts of exposed and infected cases because these are key data for public health officials, e.g., in deciding resource allocation and measures to prevent contagion.

We assume $\beta_i$ and $\beta_e$ to be equal, as precise estimates on the relative infectivity levels between the symptomatic and asymptomatic pools are unclear \cite{D2020}.
We define $\beta_{i,e}$ with decreasing piecewise constant values in time to model the escalation of the lockdown restrictions. Following the results of parameter calibration, we initially set $\beta_{i,e}=3.3\cdot 10^{-4}$ contacts$^{-1}\cdot$day$^{-1}$ on 27 February 2020, reducing this to $\beta_{i,e}=8.5\cdot 10^{-5}$ contacts$\cdot$day$^{-1}$ after the first lockdown measures on 9 March 2020, to $\beta_{i,e}=6.275\cdot 10^{-5}$ contacts$^{-1}\cdot$day$^{-1}$ after the additional restrictions on 22 March 2020, and to $\beta_{i,e}=4.125\cdot 10^{-5}$ contacts$^{-1}\cdot$day$^{-1}$ following the final restrictions on 28 March 2020. Similarly, we assume $\nu_{s,e,r}$=0.0435, 0.0198, 0.0090, and 0.0075 km$^{2}\cdot$ day$^{-1}$ over the respective phases. The Allee term $A$ is set to 1,000 individuals$\cdot$km$^{-2}$, and we fix $\nu_i=1.0\cdot 10^{-4}$ km$^{2}\cdot$ day$^{-1}$ throughout, assuming that symptomatic individuals are largely immobile.

\section{Results}\label{results}

\subsection{Forecasting the spatio-temporal dynamics of the COVID-19 pandemic in Lombardy}

Fig.~\ref{fig:outbreakDevelopment} shows the evolving spatial pattern of the COVID-19 outbreak in Lombardy, beginning with exposure in Bergamo, Brescia, Cremona and Lodi. 
The contagion moves north from Lodi into Milan via the southern suburbs and eventually reaches the city center. We note that although Lodi and Cremona are the most affected areas at the onset of the outbreak, they quickly improve and avoid the explosive growth found in Milan, Bergamo and Brescia. This is consistent with the reported data \cite{Lab24}. However, we found Cremona to be somewhat underpredicted by our model. This might be attributable to the presence of the neighboring city of Piacenza, which shares its metropolitan area with Cremona but is not included in our simulations because it belongs to the adjacent region of Emilia-Romagna.

In Fig. \ref{fig:outbreakDevelopment}, we demonstrate the remarkable qualitative agreement in the outbreak dynamics between our model forecasts and data in the three main affected areas: Milan, Bergamo, and Brescia. The $R^2$ between the model forecast and the data of infected cases are 0.997, 0.977, 0.976, and 0.998 for all Lombardy, Bergamo, Brescia, and Milan, respectively. We observe that the outbreak emerges in Milan later, where it grows more steadily, eventually becoming the most affected area in Lombardy. We also note that the lockdowns appear to have effectively halted the spread in Bergamo and Brescia. These restrictions notably reduce the spread in Milan, limiting the virus to a linear growth pattern, but fail stop it.

We observe that our simulations predict a larger number of infections than the reported data. This results from using the data for deceased cases for calibration, which is comparatively more accurate than infections (see Section 3). We obtained $R^2$= 0.972 and range-normalized RMSE=7.6\% for this subgroup. Thus, the difference in predicted and measured infections suggests a lack in the reporting of real cases, probably due to the deficiencies and difficulties of testing a significant sample of the whole living population.  However, we also remark that COVID-19 mortality data depend on the currently unknown transmission rates, which emphasizes the importance of qualitative agreement to test novel modeling approaches. To this end, we show that it is possible to rescale our simulation results to accurately match the order of magnitude of the reported infected case data Fig. 2, though we emphasize that this is purely for the purposes of visualization.

\subsection{Exploring alternative reopening scenarios}

We further use our model to assess four illustrative reopening scenarios over four months following 27 February 2020: maintenance of restrictions, relaxation of the lockdown everywhere on 3 May 2020 under two different sets of assumptions, and a combination of maintenance of restrictions in Milan and relaxation elsewhere in Lombardy.
We still consider the changes in parameter values induced by the sequential restrictions (see Section~\ref{numerics}) and the lockdown relaxation is modeled by setting $\nu_{s,e,r}=2.175\cdot 10^{-2}$ km$^2 \cdot$ day$^{-1}$ and $\beta_{i,e}=9.0\cdot 10^{-5}$ contacts$^{-1}\cdot$day$^{-1}$ (scenario A), and $\beta_{i,e}=6.6\cdot 10^{-5}$ contacts$^{-1}\cdot$day$^{-1}$ (scenario B). Scenario A is a pessimistic scenario, which assumes that the population contact rate is similar to early-outbreak levels. Scenario B is more optimistic and assumes that the generally greater public awareness of preventative measures (such as mask-wearing and social distancing) translates to greater success in limiting contact, despite increased mobility.  

\begin{figure}[t]
\includegraphics[width=\linewidth]{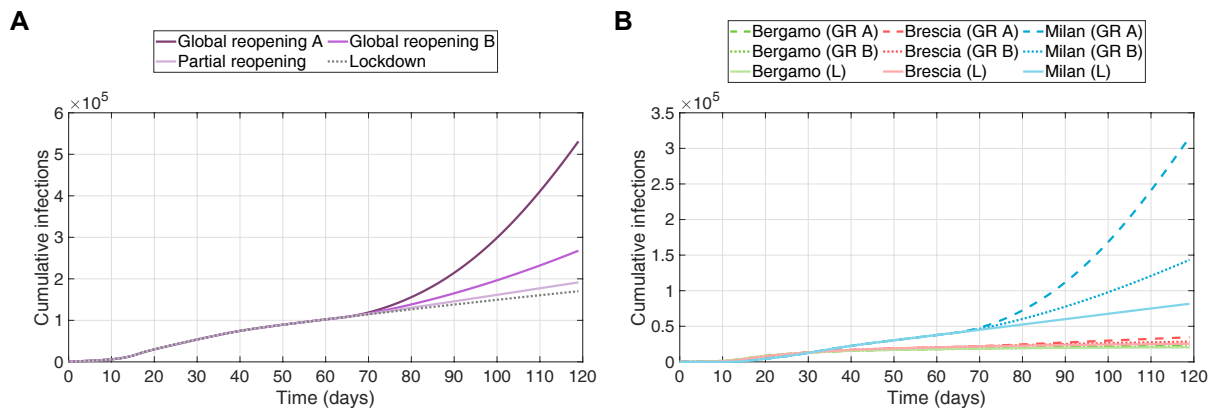}
\caption{Results of simulations over alternative reopening cases. (A) Cumulative infected cases in Lombardy over the different reopening scenarios. Our simulations suggest that maintaining a strict lockdown outside of Milan offers little benefit; however, keeping lockdown restriction in Milan may prevent explosive growth. (B) Comparison of the cumulative infections in the three largest metropolitan areas (i.e., Bergamo, Brescia and Milan) for the global reopening A (GR A, dashed lines), global reopening B (GR B, dotted lines), and maintenance of lockdown (L, solid lines).}\label{fig:reopeningComparison}
\end{figure}

Fig. \ref{fig:reopeningComparison} shows the resulting outbreak dynamics for these four reopening scenarios. Our simulations suggest that relaxing the lockdown restrictions in the entire region may cause severe and rapid growth in the Milan area. However, major urban zones far from Milan (e.g., Brescia and Bergamo) just experience a marginal increase in growth and still show a favorable trend in time.  Conversely, if we maintain the lockdown restrictions in Milan and relax them elsewhere, the outbreak shows more favorable dynamics, similar to those obtained for Brescia and Bergamo. 
Thus, our results suggest that maintenance of lockdown measures in high-population, high-density areas like Milan may be necessary for longer times to effectively arrest the spread of contagious diseases like COVID-19.

\section{Discussion}\label{disc}
We have introduced a compartmental PDE model describing spatio-temporal propagation of disease contagion and applied it to the 2020 outbreak of COVID-19 in Lombardy specifically. Our simulations are intended to provide a proof-of-concept of the potential of PDEs for regional modeling of the outbreak. Nonetheless, they show good qualitative agreement with reality, accurately predicting the outbreak dynamics in different areas and recreating the transmission path in time and space. We then used the model to examine some possible reopening scenarios, which suggested that reopening may be best determined based on local population and contagion dynamics, and not by a one-size-fits-all approach.  

Our model is in a very early stage, with ample room for improvement. We plan to consider non-constant model parameters and adaptively update them according to measured data using data-assimilation procedures \cite{GEvensen}. Indeed, as more reliable data becomes available, we can further extend parameter calibration to fit data for additional model compartments other than the deceased subgroup. Boundary conditions can also be defined in a more realistic manner, e.g., by including 0D SEIR models describing the fluxes with respect to neighboring regions. Additionally, we used population-dependent diffusion terms, but ideally these could also be affected by geographical features (e.g., rivers, mountains, roadways, and railways) \cite{KGV2013}.  Non-local effects like the ones modeled by fractional operators can also be included \cite{khan2020modeling}. These considerations may be crucial whenever using the model in larger geographical domains.

We would also like to extend our framework into more sophisticated compartmental models including, e.g., hospitalizations, patients in intensive care units, or age and biological sex structures \cite{GBB2020, Gatto202004978}. This would further increase the utility of the model, potentially helping decision-makers to determine the allocation of resources among different areas. The present results clearly pinpoint the current standpoints of virologists, emphasizing the need of restrictions. Finally, the socio-economical costs of lockdowns are not included here, but could ultimately be incorporated in future quantitative analyses, e.g., aiming at the comprehensive optimization of pandemic-arresting measures.
\section*{Acknowledgments} The authors would like to acknowledge the work of Marco Demarziani, Luigi Greco, Isabella Atcha, Kasey Cervantes, Sanne Glastra, Shreya Rana, Stefano Minelli, Chiara Macchello, Simona Petralia, Anita de Franco, and Martina Moschella for their crucial help with data acquisition.

\bibliographystyle{plain}
\bibliography{covid19_v1.bib}

\end{document}